\documentstyle{aipproc}
\input{epsf}

\begin{document}
\title{Experiments towards quantum information with trapped Calcium ions}

\author{ D. Leibfried, C. Roos$^{\dagger}$, P. Barton, H. Rohde, S. Gulde,\\ A. B. Mundt, G. Reymond$^{*}$, M. Lederbauer,  F. Schmidt-Kaler,\\ J. Eschner, and R. Blatt}
\address{Institut f\"ur Experimentalphysik, Universit\"at Innsbruck, A-6020 Innsbruck, Austria \\
$^{\dagger}$present address: Ecole Normale Superieure, Paris, France\\
$*$ present address: Institute d'Optique, Orsay, France}

\maketitle

\begin{abstract}
Ground state cooling and coherent manipulation of ions in an rf-(Paul) trap is the prerequisite for quantum information experiments with trapped ions. With resolved sideband cooling on the optical S$_{1/2}$ -D$_{5/2}$ quadrupole transition we have cooled one and two $^{40}$Ca$^+$ ions to the ground state of vibration with up to 99.9\% probability. With a novel cooling scheme utilizing electromagnetically induced transparency on the S$_{1/2}$ -P$_{1/2}$ manifold we have achieved simultaneous ground state cooling of two motional sidebands 1.7 MHz apart. Starting from the motional ground state we have demonstrated coherent quantum state manipulation on the S$_{1/2}$ -D$_{5/2}$ quadrupole transition at 729 nm. Up to 30 Rabi oscillations within 1.4 ms have been observed in the motional ground state and in the $n=1$ Fock state. 
In the linear quadrupole rf-trap with 700 kHz trap frequency along the symmetry axis (2 MHz in radial direction)  the minimum ion spacing  is more than 5 $\mu$m for up to 4 ions. We are able to cool two ions to the ground state in the trap and individually address the ions with laser pulses through a special optical addressing channel.
\end{abstract}
\section*{Introduction}
Even elementary quantum information processing operations put severe demands on the experimental techniques. A single quantum gate, for instance, requires two strongly interacting quantum systems, highly isolated from environmental disturbances. In their proposal for a quantum logic gate, Cirac and Zoller showed that single ions, trapped in a linear radio-frequency (Paul) trap and cooled to the motional ground state have the potential to offer such a system\cite{cirac95} thus initiating experimental work towards quantum logic with ion traps in several groups around the globe.   

The work done in Innsbruck is based on $^{40}$Ca$^+$ ions which offer the advantage that light sources for all transitions involved are derived from diode and solid state lasers in a relatively easy way. So far we have investigated several ways to cool one or two ions to the ground state of motion in two different traps. In the first trap, a mm-sized spherical Paul trap loaded with a single ion, we have developed and refined our cooling techniques and the coherent manipulations necessary for quantum information processing with ions. The second trap, with a linear geometry, was used to manipulate two and more ions, namely for ground state cooling and individual addressing.  

\section*{Experimental Setup}
\subsection*{Relevant Transitions in $^{40}$Ca$^+$}
There are two basic prerequisites for an ion species to be suitable for quantum information processing. First they should allow for cooling to the ground state, typically with a Doppler precooling step, and second their electronic levels should contain at least two long lived states that can form an effective two level system to encode the quantum bit (qubit). 

\begin{figure}[ht]
\epsfxsize=\hsize \centerline{\epsfbox{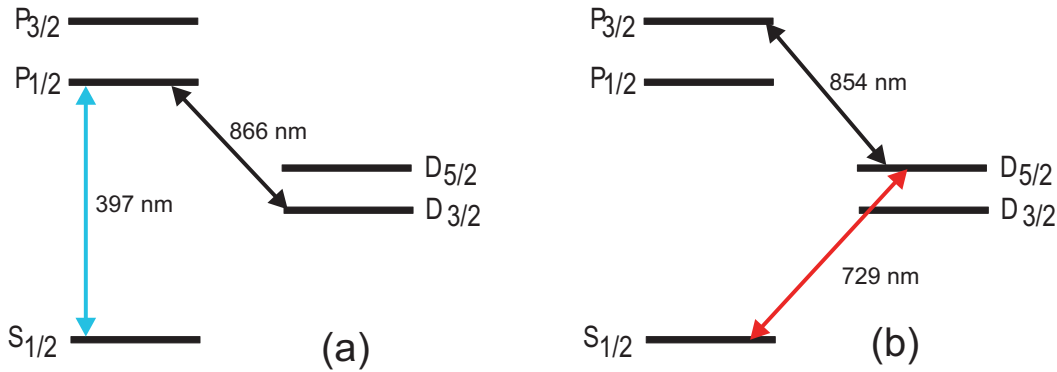}} 
\caption{\label{levscheme} Relevant levels in $^{40}$Ca$^+$ for 
(a) Doppler cooling and state detection (b) resolved sideband 
cooling and coherent manipulations are performed on the qubit 
effective two level system formed by S$_{1/2}$ and D$_{5/2}$ 
levels.} 
\end{figure}

Calcium ions can accommodate both these prerequisites with the additional benefit that all necessary light sources can be generated from diode and solid state lasers. The levels relevant to our experiments are shown schematically in Fig. \ref{levscheme}. Doppler cooling is achieved by driving the dipole transition from the S$_{1/2}$ to the P$_{1/2}$ level at 397 nm with a frequency-doubled Ti:Sapph laser that is red detuned from the transition line center by about 20 MHz, the natural decay width of the P$_{1/2}$ level. To prevent pumping to the D$_{3/2}$ level (branching ratio 16:1), the ions are simultaneously irradiated with a diode laser at 866 nm that drives the D$_{3/2}$-P$_{1/2}$ transition (Fig. \ref{levscheme} (a)). To provide a well defined direction of magnetic field and to split the Zeeman sublevels we produce a field of about 4 Gauss at the position of the ion. With an additional beam at 397 nm that is $\sigma^+$ polarized we can prepare the electronic state of the ion(s) to be S$_{1/2}$, $m=1/2$. 

For the two qubit levels we chose the S$_{1/2}$ ground state and the metastable D$_{5/2}$ state with a natural lifetime of about 1 s (Fig. \ref{levscheme}(b)). Quadrupole transitions between these levels are driven with a Ti:Sapphire laser at 729 nm, stabilized with the Pound-Drever-Hall method to a high finesse cavity. To maintain the coherence necessary for qubit manipulations this laser has to be highly stable. We have determined an upper bound of 76(5) Hz (FWHM) for the effective linewidth of our laser system by observing the fringe contrast in high resolution Ramsey spectroscopy on the S$_{1/2}, m=-1/2$-D$_{5/2}, m=-5/2$ transition as a function of the time delay between the two excitation pulses \cite{unpublished}. 

To initialize the qubit in the S$_{1/2}$ ground state we use an additional diode laser at 854 nm that repumps the ion(s) from the D$_{5/2}$ level via the P$_{3/2}$ level. All our laser sources can be switched with a timing accuracy of better than 1 $\mu$s and frequency tuned with acoustooptic modulators (AOMs). 

\subsection*{Qubit State Detection}

The laser beams at 397 nm and 866 nm used for Doppler cooling also provide highly efficient state detection with the quantum jump technique\cite{dehmelt75}. The two light fields couple the S$_{1/2}$ ground state to a cycling transition so that many photons are scattered if the ion is in the ground state. On the other hand if the ion is in D$_{5/2}$, this level is decoupled from the excitation and no fluorescence photons will be emitted. Although only a small fraction (ca. 10$^{-2}$) of these fluorescence photons is collected by a lens and imaged onto a photomultiplier with a quantum efficiency of about 10\%, one can distinguish the two qubit states within 2 ms of detection time. This allows us to measure the state of our qubit with practically 100\% efficiency.

\subsection*{Ion Traps}

In our experiments we use two different ion traps. The first trap is a regular spherical Paul trap, with ring and endcaps made of 0.2 mm diameter molybdenum wire. The ring diameter is 1.4 mm and the endcap distance 1.2 mm. With an rf-drive voltage of about 1 kV at 20 MHz we obtain motional frequencies of up to 4.5 MHz and 2 MHz along the axis of symmetry and in the ring plane respectively. This trap is mainly used for experiments with one ion. Due to its high secular frequencies, Doppler cooling leads to relatively low average occupation numbers ($\bar{n}$=2 (4) for the axial (radial) harmonic oscillator) and thus makes it a good test bed for cooling and coherent control techniques with just one ion.

Our second trap is a linear quadrupole trap with four 0.6 mm diameter quadrupole rods and the endcaps made of stainless steel. The trap is held together by Macor spacers. The diagonal distance between quadrupole electrodes is 2.36 mm and the distance between the cylindrical 6 mm diameter endcap rings is 10 mm. In this trap we reach secular frequencies of 2 MHz in the radial direction and up to 700 kHz in the axial direction. For a string of up to 4 ions this leads to inter-ion distances  $\geq$~5~$\mu$m, well above the diffraction limit of our laser beams. With this we are able to individually address such a string (see below), at the expense of relatively low axial secular frequencies and a higher average occupation number ($\bar{n}\simeq$ 25 in the axial direction) after Doppler cooling.  
\section*{Individual Addressing}

One of the key features in the Cirac-Zoller \cite{cirac95} gate is that the internal states of ions in a string have to be manipulated individually. Although a number of technically different proposals have been made, the original approach of Cirac and Zoller, to focus a laser beam sufficiently so it will only interact with a single ion in the string, still seems to be the most straightforward way. One obvious limitation of this approach is that the size of the focus is limited by diffraction to roughly a micron so the minimum distance between ions has to be larger than that number. A given minimum spacing of ions restricts the maximum center of mass (COM) frequency for a given number of ions along the axis of symmetry \cite{steane97}. If four ions of Ca$^+$ should not be closer together than 5 $\mu$m, the maximum COM frequency is about 700 kHz. 

\begin{figure}[ht]
\epsfxsize= \hsize \centerline{\epsfbox{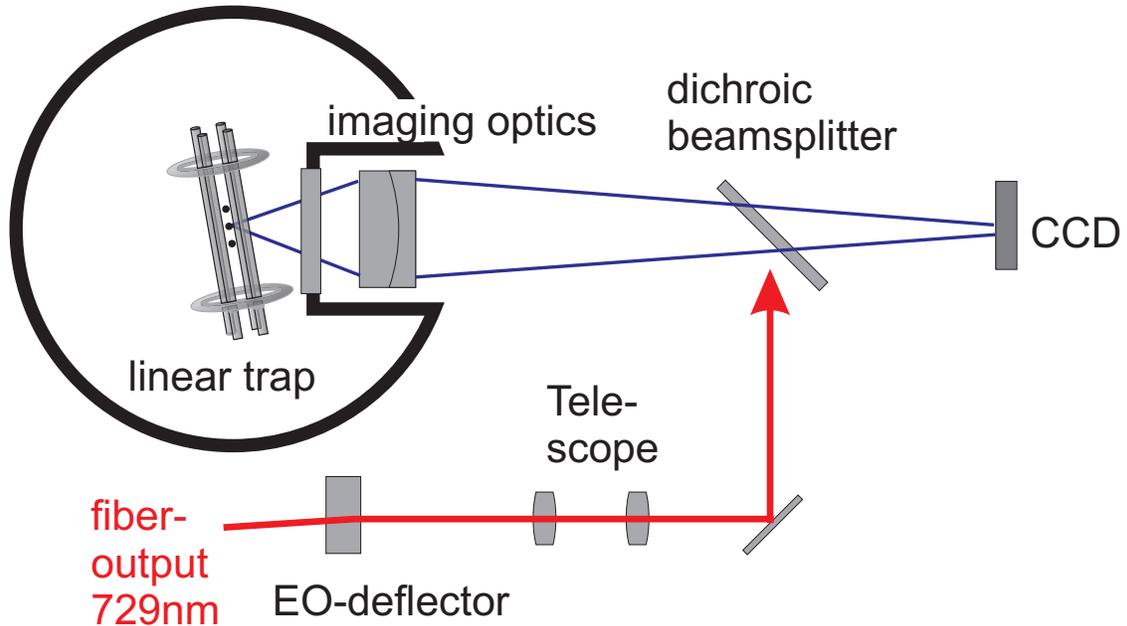}} 
\caption{\label{indadr} Setup for individual addressing of a 
string of ions. The  addressing beam is displaced by an 
electrooptic (EO) deflector and superimposed onto the CCD imaging 
channel with a dichroic beamsplitter. By sending it through the 
imaging lens system in reverse we utilize the high spatial 
resolution of this system for addressing.} 
\end{figure}

To image the fluorescence of ion strings, we use a Nikon MNH-23150-ED-Plan-1,5x macroscope lens with a working distance of 65 mm and an intensified CCD camera (see figure \ref{indadr})\cite{naegerl98a}. The spatial resolution of our imaging system is about 2 $\mu$m, sufficient to resolve single ions in not too densely packed strings. For individual addressing we use the imaging lens in reverse. 
Since the addressing beam is at 729 nm we can use a dichroic mirror to superimpose it with the imaging channel (see figure \ref{indadr}). A telescope is used to sufficiently expand the beam diameter \cite{naegerl99a}. The beam is steered over the ions with an electrooptic deflector. The deflection efficiency of 5 mrad/kV is translated into a displacement on the ions of 23 $\mu$m/kV. A high voltage amplifier stage for addressing allows us to switch from one ion to the other in a few $\mu$s. We have checked the beam diameter and pointing stability of our system by mapping the Rabi frequency on the S$_{1/2}$-D$_{5/2}$ transition versus the beam displacement and found a 1/e width of 3.7(0.3) $\mu$m for this excitation. If we apply a $\pi$-pulse to the ion addressed, the probability of exciting a neighboring ion in the ground state and 5 $\mu$m away would be about 1\%.

\section*{Ground State Cooling}
\subsection*{Resolved Sideband Cooling}
\begin{figure}[htp]
\epsfxsize=\hsize \centerline{\epsfbox{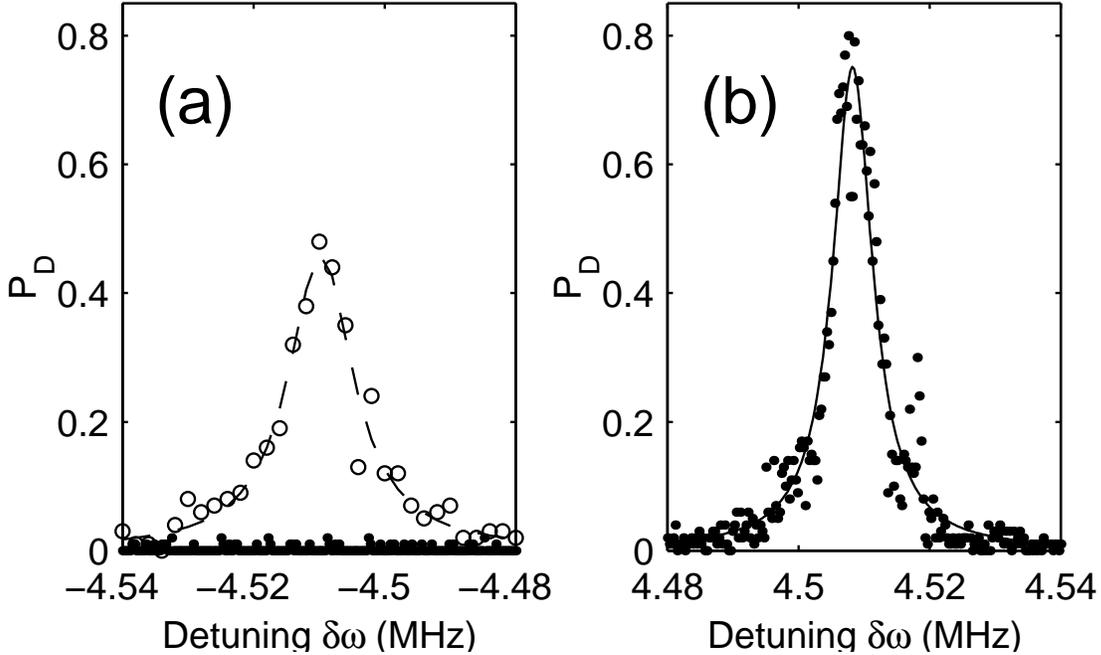}} \vspace{3 pt} 
\caption{\label{sbcooling} Red and blue sidebands of the axial 
mode (at 4.51 MHZ) of one ion after Doppler cooling (dashed) and 
after resolved sideband cooling (solid). From the ratio of 
sideband strengths we determine 99.9\% ground state occupation.} 
\end{figure}

Ground state cooling has been achieved so far with a single $^{199}$Hg$^+$
ion \cite{diedrich89}, and with $^9$Be$^+$ \cite{monroe95a}, using resolved sideband cooling on either a quadrupole or a Raman transition. We use a cooling  method on the S$_{1/2}$-D$_{5/2}$ quadrupole transition, similar to the Hg experiment \cite{roos99}. The weak coupling between light and atom on a bare quadrupole transition would necessitate long cooling times. However, the cooling rate is greatly enhanced by (i) strongly saturating the transition and (ii) shortening the lifetime of the excited state via coupling it to a dipole-allowed transition. The $S_{1/2}(m=1/2)-D_{5/2}(m=5/2)$ transition, well resolved in frequency from all other possible transitions by the applied magnetic field, is excited with about 1 mW of light focused to a waist size of 30 $\mu m$ at the position of the ion. At the same time, the decay rate back to the ground state is increased by exciting the $D_{5/2}(m=5/2)-P_{3/2}(m=3/2)$ transition. The intensity of this quenching laser is adjusted for optimum cooling during the experiment. Optical pumping to the $S_{1/2}(m=-1/2)$ level is prevented by occasional short laser pulses of $\sigma^+$ polarized light at 397 nm. The duration of those pulses is kept at a minimum to prevent unwanted heating. The ground state occupation is found by comparison of the on resonance excitation probability for red and blue sideband transitions \cite{roos99}. In the spherical Paul-trap we reach up to 99.9$\%$ of motional ground state occupation within 6 ms (see figure \ref{sbcooling}). 
All earlier successful ground state cooling experiments were plagued by an unexpectedly high motional heating (see \cite{turchette00} and references therein). In our setup we find a motional heating rate of one phonon in 190 ms for a trap frequency of 4 MHz, two orders of magnitude smaller than in the trap used at NIST for the $^9$Be$^+$ experiment. While this is still a much higher heating rate than expected from black body radiation it happens on a timescale much longer than the time typically needed for quantum logic gates (we estimate an upper limit of 200 $\mu$s for one CNOT gate). 

In separate experiments we also cooled all motional modes of two ions in our linear trap to the ground state \cite{schmidt00}. Cooling only one of the 6 motional modes at a time we achieved at least 95\% ground state occupation for all modes. In these experiments we used the addressing channel to illuminate only one ion with the cooling radiation. The second ion is cooled sympathetically due to the strong inter-ion coupling by the Coulomb-force. 
\subsection*{Cooling with Electromagnetically Induced Transparency}
Resolved sideband cooling only leads to very low temperatures if the red sideband is excited with a narrow excitation bandwidth. Otherwise nearby nonresonant transitions (e.g. carrier transitions) will lead to unwanted excess heating and severely increase the final temperature of the ion(s). Unless sideband frequencies are degenreate this limits resolved sideband cooling to one motional sideband at a time. For more degrees of freedom, for example the 6 motional modes of two ions, resolved sideband cooling needs involved cooling schemes where all degrees of freedom have to be cooled sequentially with laser detuning and laser power appropriate for their coupling strength. Moreover the  phonons scattered in the process of cooling one motional mode will reheat the other modes. 

For a Cirac-Zoller gate only the motional mode that is used as the 'quantum-bus' has to be cooled to very high ground state occupation as long as the other motional modes are inside the Lamb-Dicke regime \cite{wineland98}. This regime can be reached in principle by Doppler cooling, as long as the trap frequencies are not too different from the natural linewidth of the cooling transition. Therefore typically trap frequencies on the order of 10 MHz or higher are needed to reach the Lamb-Dicke regime, but will result in an ion spacing that is hard to optically resolve. This leads to a problematic situation: either one has difficulties in addressing the ions or in sufficiently cooling all vibrational modes of a string. In our experiments with two or more ions in the linear trap we decided to maintain good conditions for individual addressing and limited our axial trap COM frequency to 700 kHz or lower.   

\begin{figure}[htp]
\epsfxsize=\hsize \centerline{\epsfbox{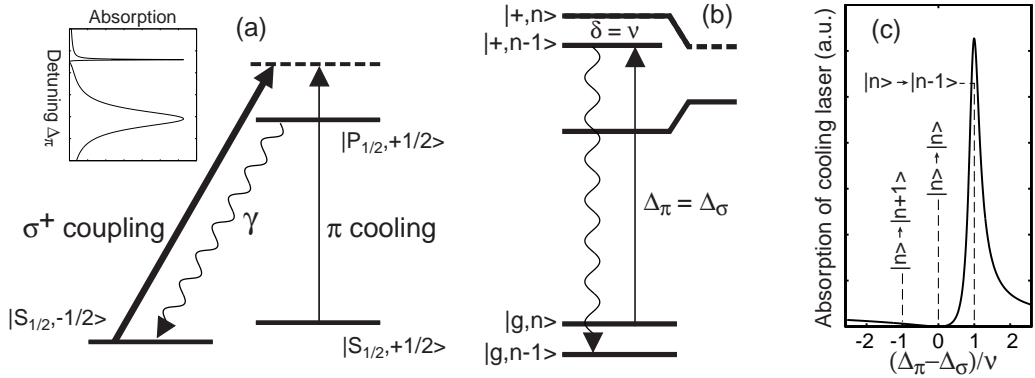}} 
\caption{\label{eitscheme} (a) Level scheme of the 
S$_{1/2}$-P$_{1/2}$ manifold used for EIT cooling. The bare levels 
are dressed with a strong $\sigma^+$ polarized beam. This causes a 
distinct asymmetry in the absorption profile of a weak 
$\pi$-polarized beam. (b) Dressed state picture: with light shift 
$\delta$ = trap frequency $\nu$, the cooling laser is resonant 
with the transition $|g,n\rangle \to |+,n-1\rangle$ where 
$|g\rangle$ is the S-state and $|+\rangle$ is the dressed state 
corresponding to the virtual level [dashed line in (a)]. (c) By 
detuning the $\pi$-polarized beam to the carrier transition 
frequency and choosing the AC-Stark shift caused by the dressing 
beam to be equal to the trap frequency, the asymmetry of the 
absorption profile is superimposed over the carrier and sideband 
transitions in such a way that the cooling red sideband 
transitions are much more probable as compared to the heating blue 
sideband transitions, and the carrier is completely suppressed by 
a dark resonance.} 
\end{figure}
Under these conditions it was desirable to find a cooling technique that is not as narrow-band as resolved sideband cooling but has a lower cooling limit than Doppler cooling. Ideally one would want to cool the ion deeply into the Lamb-Dicke regime for all motional degrees of freedom simultaneously. A very recent proposal to use electromagnetically induced transparency (EIT) for the cooling of trapped particles \cite{morigi00} held this promise.

We adapted this cooling scheme, originally proposed for a three-level system, for the case of the [S$_{1/2}$, P$_{1/2}$] four level system in Ca$^+$ that we also use for Doppler cooling. The manifold is dressed with a $\sigma^+$ polarized beam at 397 nm, blue detuned by $\Delta_{\sigma}$=60 MHz (3 linewidths of the S-P transition) that connects the S$_{1/2}$, m=-1/2 with the P$_{1/2}$, m=1/2 level. Under these circumstances a second low intensity $\pi$ polarized beam will experience an absorption (Fano-) profile as depicted in Fig. \ref{eitscheme}(a). In addition to the usual line profile around $\Delta_{\pi}=0$ a dark resonance is created at the point where the detuning of the $\pi$-plarized beam $\Delta_{\pi}$ is equal to $\Delta_{\sigma}$ and a bright resonance appears at $\Delta_{\pi}=\Delta_{\sigma}+\delta$ where $\delta$ is the AC Stark shift due to the $\sigma^+$ polarized beam. This creates an asymmetry in absorption for carrier and sidebands. The carrier is almost completely suppressed due to the dark resonance, the blue sideband is in the shallow wing of the profile, but the red sideband is greatly enhanced by the bright resonance created by the dressing beam.

For cooling the $\pi$-polarized beam is tuned to the {\it carrier}. Absorption on the carrier is then suppressed by the dark resonance while absorption on the red sideband is enhanced by the bright resonance. When we tuned the Stark-shift $\delta$ to be equal to one of the motional modes at 3.34 MHz we were able to cool this mode to 90\% ground state occupation or $\bar{n}=0.1$.  

As sketched in figure \ref{eitscheme} (b) the bright resonance can have a substantial width and the red sideband has not necessarily to coincide exactly with the maximum of the bright resonance to get a cooling effect. This opens the possibility to cool several motional modes {\it simultaneously}, as long as they are not too far apart in frequency. 
\begin{figure}[htp]
\epsfxsize= 12 cm \centerline{\epsfbox{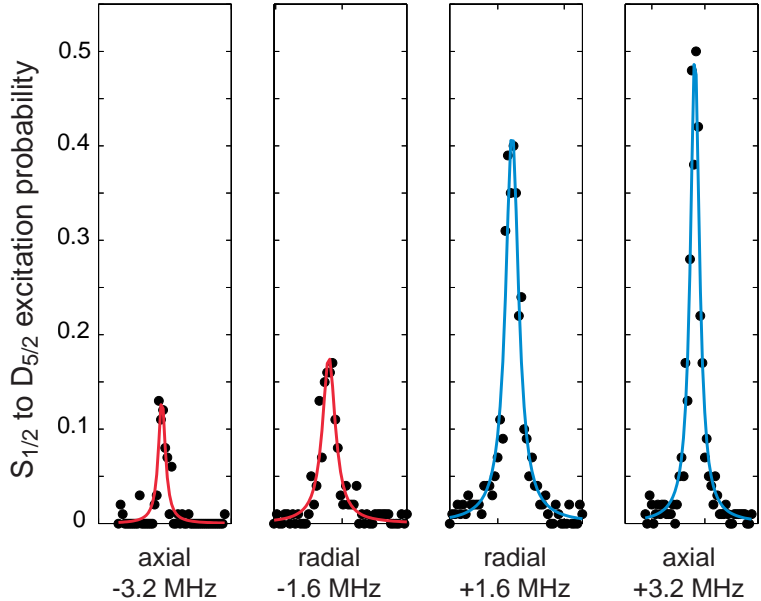}} 
\caption{\label{eitcool} Simultaneous cooling of two vibrational 
modes with an oscillation frequency difference of 1.73 MHz. The 
sideband asymmetry corresponds to 58\% ground state occupation 
($\bar{n}=0.85$) in the mode at 1.61 MHz and 74\% ($\bar{n}=0.35$) 
at 3.34 MHz.} 
\end{figure}
To demonstrate simultaneous cooling of two vibrational modes with this method we chose two vibrational modes at 1.61 MHz and 3.34 MHz with an oscillation frequency difference of 1.73 MHz. The AC-Stark shift $\delta$ of the $\sigma^+$- beam was adjusted to be about 2.5 MHz, halfway in between the two mode frequencies. With this settings we achieved 58\% ground state occupation ($\bar{n}=0.85$) in the mode at 1.61 MHz and 74\% ($\bar{n}=0.35$) at 3.34 MHz.
From this result we estimate that we can sufficiently cool all axial degrees of freedom of a string of up to 5 ions with a COM mode frequency of 700 kHz. 
\section*{Coherent Manipulations}
For quantum information processing, it is important to know for how long coherent interaction with the ion(s) is possible. To this end we cooled one ion to the ground state and then irradiated the ion with light at the blue sideband frequency (This interaction is used in a quantum gate to transfer the internal state of a qubit into the motion)\cite{roos99}. We then monitored the occupation probability of the D-state versus the pulse length on the blue sideband. The same interaction was also used after preparing the ion in the $n=1$ motional Fock state by a $\pi$-pulse on the blue sideband followed by a repumping pulse on the D$_{5/2}$-P$_{3/2}$ transition. 
\begin{figure}[ht]
\epsfxsize=\hsize 

\centerline{\epsfbox{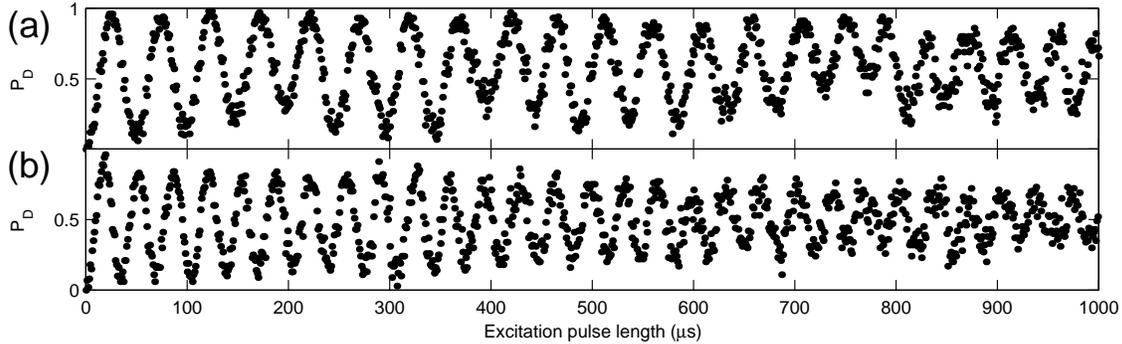}} \caption{\label{flops} (a) 
Rabi-oscillations on the blue sideband for the initial motional 
state $|n=0 \rangle$. (b) Rabi-oscillations as in (a), but for 
$|n=1\rangle$.} 
\end{figure}
As figure \ref{flops} shows, we were able to observe Rabi-flops for both initial motional states with a contrast of better than 50\%  for 1 ms. The ratio of Rabi-frequencies is $\sqrt{2}$ as expected for this kind of interaction in the Lamb-Dicke regime. These results make us confident that we should be able to apply gate pulses equivalent to at least 40 $\pi$-pulses before the fidelity of the total operation drops below 0.5. In our system motional heating is too slow to be the prime source of the observed decoherence, we rather attribute it to time dependent magnetic field fluctuations that shift the levels and to slow vibrations in our setup that introduce a fluctuating Doppler-shift for our optical beam at 729 nm. The linewidth of our laser system could also contribute, but from the results of the Ramsey-experiment described above we would expect this source to contribute with a characteristic time of about 15 ms.
%
%
\section*{Conclusions and Outlook}
\begin{figure}[ht]
\epsfxsize=\hsize \centerline{\epsfbox{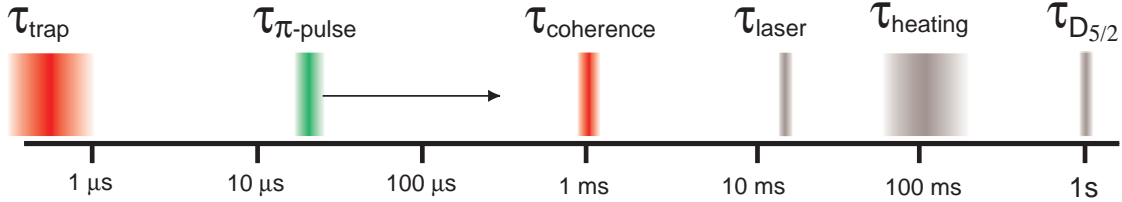}} 
\caption{\label{scales} Characteristic time scales for our 
experimental setup. For more details see text.} 
\end{figure}

In conclusion we have demonstrated that we have all necessary ingredients to perform a two bit quantum logic gate with trapped ions. The time scales in our system are well resolved (see figure \ref{scales}).
The fastest characteristic time is given by the harmonic motion of the ion(s) and on the order of 1 $\mu$s. A $\pi$-pulse on a motional sideband takes about 20 $\mu$s and, as stated above, the fidelity of coherent manipulations remains above 0.5 for times smaller than 1 ms. Our laser system would allow for coherent manipulations for at least 15 ms. Motional heating begins to play a role for times around 100 ms. The ultimate source of decoherence in our experiment is the 1 s lifetime of the D$_{5/2}$ state.

In the near future we plan to use our ability to individually address ions to demonstrate a CNOT quantum logic gate with two ions and to create maximally entangled states with 2 and more ions. We also envision to perform small quantum algorithms and first experiments on error corrections with up to 5 ions.

{\bf Acknowledgements}
This work was supported by the Fonds zur F\"orderung wissenschaftlicher Forschung (FWF) within the special research grant SFB 15, by the European Commission within the TMR networks 'Quantum Information' (ERB-FMRX-CT96-0087) and 'Quantum Structures' (ERB-FMRX-CT96-0077) and by the Institut f\"ur Quanteninformation GmBH. 
\newpage

\end{document}